\documentclass[aps,pre,floats,twocolumn,showpacs,superscriptaddress]{revtex4-1}

\usepackage{graphics,dcolumn,bm,epic, eepic,fleqn,float}
\usepackage{graphicx}
\usepackage{dcolumn}
\usepackage{bm}
\usepackage{epsfig}
\usepackage{color}
\usepackage{natbib}

\usepackage{amssymb,amsmath,amsfonts,multirow,rotate,xcolor}
\usepackage{subfigure}
\usepackage{bbm}

\newcommand{\lay}[1]{^{[#1]}}
\newcommand{\avg}[1]{\langle #1 \rangle}
\usepackage{ulem}

\newcommand{\Tx}[1]{}

\begin{document}

\title{The multiplex dependency structure of financial markets}

\author{Nicol\'o Musmeci}
\affiliation{Department of Mathematics, King's College London, The Strand, London WC2R 2LS, UK}
\author{Vincenzo Nicosia}
\affiliation{School of Mathematical Sciences, Queen Mary, University of London, Mile End Road, London E1 4NS, UK }
\author{Tomaso Aste}
\affiliation{Department of Computer Science, University College London, Gower Street, London, WC1E 6BT, UK }
\affiliation{Systemic Risk Centre, London School of Economics and Political Sciences, London, WC2A 2AE, UK}
\author{Tiziana Di Matteo}
\affiliation{Department of Mathematics, King's College London, The Strand, London WC2R 2LS, UK}
\affiliation{Department of Computer Science, University College London, Gower Street, London, WC1E 6BT, UK }
\author{Vito Latora}
\email{correspondence to be sent to: v.latora@qmul.ac.uk}
\affiliation{School of Mathematical Sciences, Queen Mary, University of London, Mile End Road, London E1 4NS, UK }

\begin{abstract}
We propose here a multiplex network approach to investigate
simultaneously different types of dependency in complex data sets. In
particular, we consider multiplex networks made of four layers
corresponding respectively to linear, non-linear, tail, and partial
correlations among a set of financial time series. We construct the
sparse graph on each layer using a standard network filtering
procedure, and we then analyse the structural properties of the
obtained multiplex networks.  The study of the time evolution of the
multiplex constructed from financial data uncovers important changes
in intrinsically multiplex properties of the network, and such changes
are associated with periods of financial stress. We observe that some
features are unique to the multiplex structure and would not be
visible otherwise by the separate analysis of the single-layer
networks corresponding to each dependency measure.
\end{abstract}

\flushbottom \maketitle

\section*{Introduction}
In the last decade network theory has been extensively applied to the
analysis of financial markets.  Financial markets and complex systems
in general are comprised of many interacting elements, and
understanding their dependency structure and its evolution with time
is essential to capture the collective behaviour of these systems, to
identify the emergence of critical states, and to mitigate systemic
risk arising from the simultaneous movement of several
factors. Network filtering is a powerful instrument to associate a
sparse network to a high-dimensional dependency measure and the
analysis of the structure of such a network can uncover important
insights on the collective properties of the underlying system.
Following the line first traced by the preliminary work of Mantegna
\cite{mantegna1}, a set of time series associated with financial asset
values is mapped into a sparse complex network whose nodes are the
assets and whose weighted links represent the dependencies between the
corresponding time series.  Filtering correlation matrices has been
proven to be very useful for the study and characterization of the
underlying interdependency structure of complex datasets
\cite{mantegna1,Tumminello05,PMFG2,PMFG3,exploring_genus}. Indeed, sparsity
allows to filter out noise, and sparse networks can then be analyzed
by using standard tools and indicators proposed in complex networks
theory to investigate the multivariate properties of the dataset
\cite{Boccaletti2006,TMFG}.  Further, the filtered network can be used
as a sparse inference structure to construct meaningful and
computationally efficient predictive models~\cite{TMFG,LoGo}.

Complex systems are often characterized by non-linear forms of
dependency between the variables, which are hard to capture with a
single measure and are hard to map into a single filtered network. A
multiplex network approach, which considers the multi-layer structure
of a system in a consistent way, is thus a natural and powerful way to
take into account simultaneously several distinct kinds of dependency.
Dependencies among financial time series can be described by means of
different measures, each one having its own advantages and drawbacks,
and this has lead to the study of different type of networks, namely
correlation networks, causality networks, etc. The most common
approach uses Pearson correlation coefficient to define the weight of
a link, because this is a quantity that can be easily and quickly
computed. However, the Pearson coefficient measures the linear
correlation between two time series \cite{pitfalls_corr}, and this is
quite a severe limitation, since nonlinearity has been shown to be an
important feature of financial markets \cite{sornette_nonlinear}.
Other measures can provide equally informative pictures on assets
relationships.  For instance, the Kendall correlation coefficient
takes into account monotonic non-linearity \cite{kendall}
\cite{Meissner}, while others measures, such as the tail dependence,
quantify dependence in extreme events.  It is therefore important to
describe quantitatively how these alternative descriptions relate but
also differ from the Pearson correlation coefficient, and also to
monitor how these differences change in time, if at all.

In this work we exploit the power of a multiplex approach to analyse
simultaneously different kinds of dependencies among financial time
series. The theory of multiplex network is a recently introduced
framework that allows to describe real-world complex systems
consisting of units connected by relationships of different kinds as
networks with many layers, and where the links at each layer represent
a different type of interaction between the same set of
nodes~\cite{multiplex_review,metrics_multiplex}.  A multiplex network
approach, combined with network filtering, is the ideal framework to
investigate the interplay between linear, non-linear and tail
dependencies, as it is specifically designed to take into account the
peculiarity of the patterns of connections at each of the layers, but
also to describe the intricate relations between the different
layers~\cite{Nicosia2014correlations}.

The idea of analyzing multiple layers of interaction was introduced
initially in the context of social networks, within the theory of
frame analysis \cite{Goffman}. The importance of considering multiple
types of human interactions has been more recently demonstrated in
different social networks, from terrorist organizations
\cite{metrics_multiplex} to online communities; in all these cases,
multilayer analyses unveil a rich topological structure \cite{Szell},
outperforming single-layer analyses in terms of network modeling and
prediction as well \cite{Klimek}\cite{Corominas-Murtra}. In particular
multilayer community detection in social networks has been shown to be
more effective than single-layer approaches \cite{Tang}; similar
results have been reported for community detection on the World Wide
Web \cite{Kolda}\cite{Barnett} and citation networks \cite{Wu}.  For
instance, in the context of electrical power grids, multilayer
analyses have provided important insight into the role of
synchronisation in triggering cascading failures
\cite{Buldyrev}\cite{Brummitt}. Similarly, the analyses on transport
networks have highlighted the importance of a multilayer approach to
optimise the system against nodes failures, such as flights
cancellation \cite{Cardillo}. In the context of economic networks,
multiplex analyses have been applied to study the World Trade Web
\cite{Barigozzi}.  Moreover, they have been extensively used in the
context of systemic risk, where graphs are used to model interbank and
credit networks \cite{Montagna}\cite{Burkholz}.  

Here, we extend the multiplex approach to financial market time
series, with the purpose of analysing the role of different measures
of dependences namely the Pearson, Kendall, Tail and Partial
correlation.  In particular we consider the so-called Planar Maximally
Filtered Graph (PMFG) \cite{Tumminello05} \cite{PMFG2} \cite{PMFG3} \cite{TMFG} as
filtering procedure to each of the four layers. For each of the four
unfiltered dependence matrices, the PMFG filtering starts from the
fully connected graph and uses a greedy procedure to obtain a planar
graph that connects all the nodes and has the largest sum of
weights~\cite{PMFG2} \cite{PMFG3}. The PMFG is able to retain a higher number of
links, and therefore a larger amount of information, than the Minimum
Spanning Tree (MST) and can be seen as a generalization of the latter
which is always contained as a proper
sub-graph~\cite{Tumminello05}. The topological structures of MST and
PMFG have been shown to provide meaningful economic and financial
information~\cite{mst_exch_rate,pozzi_dyn_net,NJP10,market_mode,musmeci_DBHT}
that can be exploited for risk monitoring
\cite{black_monday,musmeci_jntf,musmeci_submitted} and asset
allocation \cite{cluster_portfolio,invest_periph}.  The advantage of
adopting a filtering procedure is not only in the reduction of noise
and dimensionality but more importantly in the possibility to generate
of sparse networks, as sparsity is a requirement for most of the
multiplex network measures that will be used in this
paper~\cite{metrics_multiplex}.  It is worth mentioning that the
filtering of the partial correlation layer requires an adaptation of
the PMFG algorithm to deal with asymmetric relations. We have followed
the approach suggested in~\cite{partial_corr}, that rules out double
links between nodes. The obtained planar graph corresponding to
partial correlations has been then converted into an undirected graph,
and included in the multiplex.

\section*{Results}
\subsection*{Multiplex network of financial stocks}

We have constructed a time-varying multiplex network with $M=4$ layers
and a varying number of nodes. Nodes represent stocks, selected from a
data set of $N_{tot}=1004$ US stocks which have appeared at least once
in S\&P500 in the period between 03/01/1993 and 26/02/2015.  The
period under study has been divided into $200$ rolling time windows,
each of $\theta = 1000$ trading days.  The network at time
{$T=1,2,\ldots, 200$} can be described by the adjacency matrix
$a^{\alpha}_{ij}(T)$, with $i,j=1,\ldots,N(T)$ and
$\alpha=1,2,3,4$. The network at time window $T$ has $N(T) < N_{tot}$
nodes, representing those stocks which were continuously traded in
time window $T$.  The links at each of the four layers are constructed
by means of the PMFG procedure from Pearson, Kendall, Tail and Partial
dependencies (see Materials and Methods for details).

\begin{figure*}[ht!]
\includegraphics[width=6.1in]{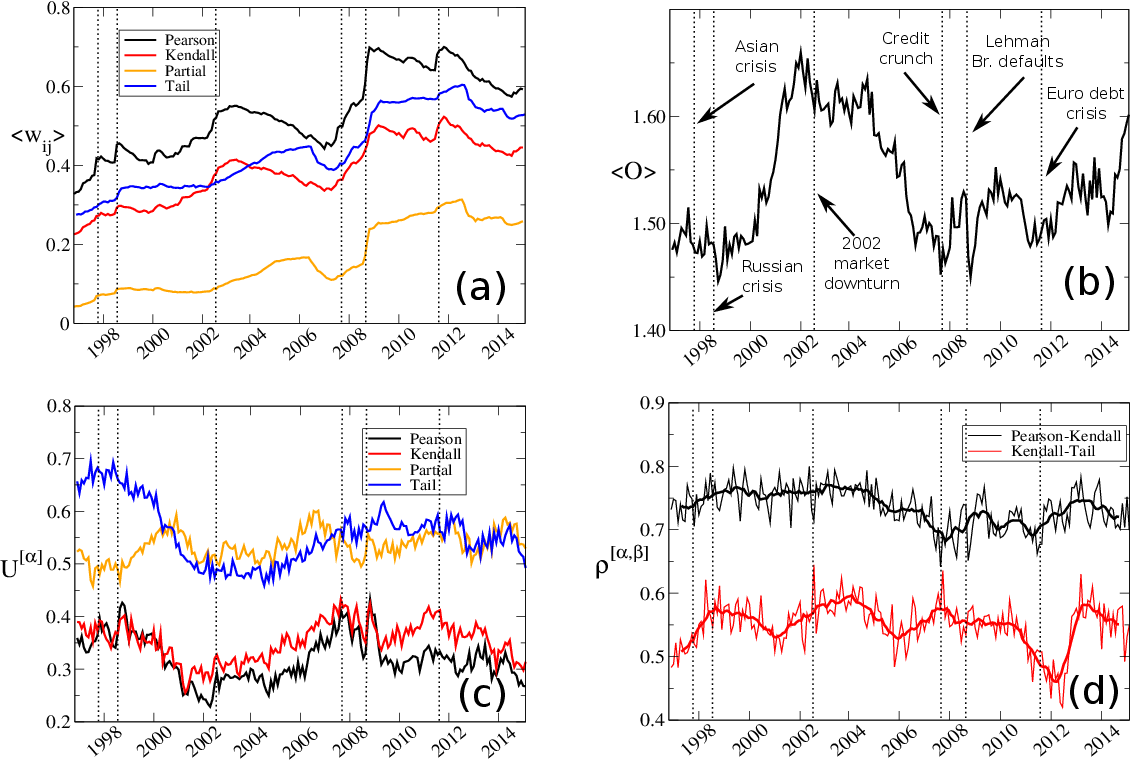}
 \caption{\label{fig:fig1}{\textbf{The multiplex nature of dependence
       among financial assets.}}  The plots report the network
   analysis of a multiplex whose four layers are Planar Maximally
   Filtered Graphs (PMFGs) obtained from four classical dependence
   measures, namely Pearson, Kendall, Tail and Partial correlation,
   computed on rolling time windows of 23 trading days between 1993
   and 2015. Each of the four layers provides different information on
   the dependency structure of a market.  Although market events and
   trends have a somehow similar effect on the average dependence
   $\langle w_{ij} \rangle$ between nodes at the different layers
   (panel (a)), each layer has a distinct local structure. This is
   made evident by the plots of the average edge overlap $\langle O
   \rangle$ (panel (b)) and of the fraction $U\lay{\alpha}$ of edges
   unique to each layer, an edge exists on average on less than two
   layers, and up to $70\%$ of the edges of a layer are not present on
   any other layer. Moreover, the same node can have different degrees
   across the four layers, as indicated by the relatively low values
   of the pairwise inter-layer degree correlation coefficient
   $\rho\lay{\alpha,\beta}$ reported in panel (d) for two pairs of
   layers over the whole observation interval. }
\end{figure*}

Fig.~\ref{fig:fig1}(a) shows how the average link weight of each of
the four dependency networks changes over time. These results indicate
an overall increase of the typical weights in the examined period
1993-2015 and reveals a strongly correlated behaviour
of the four curves (with  linear correlation coefficients between the curves range in $[0.91,0.99]$).
This strong correlation in the
temporal patterns of the four measures of dependence may lead to the
wrong conclusion that the four networks carry very similar information
about the structure of financial systems. Conversely, we shall see that even basic
multiplex measures  suggest otherwise.  In
Fig.~\ref{fig:fig1}(b) we report the average edge overlap $\langle O
\rangle$, that is the average number of layers of the financial
multiplex network where a generic pair of nodes $(i,j)$ is connected
by an edge (see Materials and Methods for details). Since our
multiplex network consists of four layer, $\langle O \rangle$ takes
values in $[1,4]$, and in particular we have $\langle O \rangle=1$
when each edge is present only in one layer, while $\langle O
\rangle=4$ when the four networks are identical. 
The relatively low values of $\langle O \rangle$ observed in this case reveal the complementary role played by the different dependency indicators. 
It is interesting to note that the edge overlap $\langle O \rangle$
displays a quite dynamic pattern, and its variations seem to be
related to the main financial crises highlighted by the vertical lines
in Fig.~\ref{fig:fig1}(b). The first event that triggers a sensible
decrease in the average edge overlap is the Russian crisis in 1998,
which corresponds to the overall global minimum of $\langle O \rangle$
in the considered interval. Then, $\langle O \rangle$ starts
increasing towards the end of year 2000 and reaches its global maximum
at the beginning of 2002, just before the market downturn of the same
year. We observe a marked decrease in 2005, in correspondence with the
second phase of the housing bubble, which culminates in the dip
associated to the credit crunch at the end of 2007. A second, even
steeper drop occurs during the Lehman Brothers default of 2008. After
that, the signal appears more stable and weakly increasing, especially
towards the end of 2014. Since each edge is present, on average, in less than 
two layers, each
of the four layers effectively provides a partial perspective on the
dependency structure of the market. This fact is made more evident by
the results reported in Fig.~\ref{fig:fig1}(c), where we show, for
each layer $\alpha=1,\ldots, 4$, the fraction of edges $U\lay{\alpha}$
that exist exclusively in that layer (see Materials and Methods for
details). We notice that, at any point in time, from $30\%$ to $70\%$
of the edges of each of the four layers are unique to that layer,
meaning that a large fraction of the dependence relations captured by
a given measure are not captured by the other measures. 

Another remarkable finding is that also the relative importance of a stock in
the network, measured for instance by its centrality in terms of
degree \cite{graph_theory,invest_periph}, varies a lot across layers. This
is confirmed by the degree correlation coefficient $\rho\lay{\alpha,
  \beta}$ for pairs of layers $\alpha$ and $\beta$.  
In general, high values of $\rho\lay{\alpha, \beta}$ signal the
presence of strong correlations between the degree of the same node in
the two layers (see Materials and Methods for details). 
Fig.~\ref{fig:fig1}(d) shows $\rho\lay{\alpha, \beta}$ as a function of time for two 
pairs of dependence measures, namely Pearson--Kendall and Kendall--Tail. Notice that 
the degrees of the layers corresponding to Pearson and Kendall exhibit 
a relatively large correlation, which remains quite stable over the
whole time interval. Conversely, the degrees of nodes in the Kendall
and Tail layers are on average less correlated, and the corresponding
values of $\rho\lay{\alpha,\beta}$ exhibit larger
fluctuations. For example, in the tenth time window we find that General Electric stock (GE US) is a hub in Kendall layer with 71 connections, but it has only 16 connections in the Tail layer: therefore the relevance of this stock in the dependence structure depends sensitively on the layer.

The presence of temporal fluctuations in $\langle O \rangle$, in
particular the fact that $\langle O \rangle$ reaches lower values during financial crises, together with the unique patterns of links 
at each layer, testified by high values of $U\lay{\alpha}$ and by 
relatively weak inter-layer degree-degree correlations for 
some pairs of layers, confirm that an analysis of relations among stocks simply 
based on one dependence measure can neglet 
relevant information which can  however be captured by other measures. 
As we will show below, a multiplex network approach which takes into account 
at the same time all the four dependence measures, but without aggregating them 
into a single-layer network, is able to provide a richer 
description of financial markets.

\begin{figure*}[ht!]
   \includegraphics[scale=0.5]{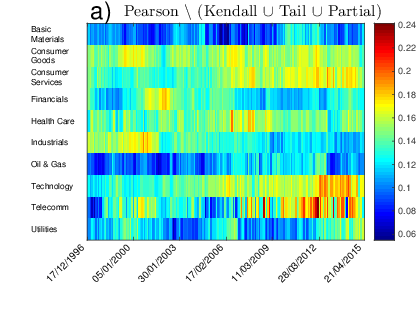}
  \includegraphics[scale=0.5]{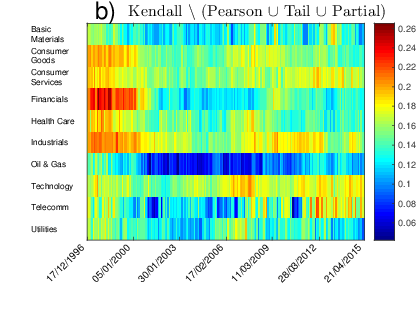}
   \includegraphics[scale=0.5]{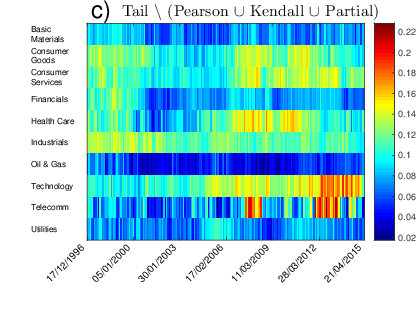}
  \includegraphics[scale=0.5]{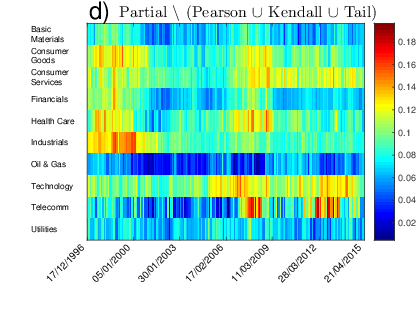}
  \includegraphics[scale=0.5]{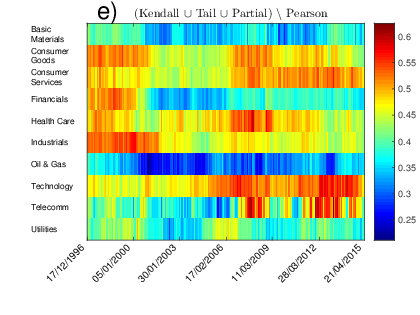}
\caption{\label{fig:multidegree} {\textbf{Multidegrees reveal the different role of industrial sectors during crises}}. 
The plots of the average multidegree of the nodes of the same industrial sector restricted to edges existing exclusively on the (a) Pearson,
  (b) Kendall, (c) Tail, and (d) Partial layers, clearly show that
  some dependence measures can reveal structures which are unnoticed
  by other measures. In particular, the plot of the average
  multidegree associated to edges existing on at least one layer among
  Kendall, Tail and Partial, but not on Pearson (panel (e)), reveals  that Pearson correlation does not capture many important features such as the prominent role of Basic Materials, Financial, Consumer Goods and Industrials during crises and the increasing importance of Technology and Consumer Services after the 2007-2008 crisis.
 }
\end{figure*}  

\subsection*{Multiedges and node multidegrees.}

As a first example of useful quantities that can be investigated in a
multiplex network, we have computed the so-called multidegree
$k_i^{\vec m}$ for each node $i$ in the network, corresponding to
different multiedges (see Materials and Methods)~\cite{Bianconi2013}.
In particular, we have normalised the multidegree of node $i$ dividing
it by the corresponding node overlapping degree $o_i$, so that the
resulting $k_i^{\vec m}/o_i$ is the fraction of multiplex edges of
node $i$ that exist only on a given subset of layers.  In
Fig.~\ref{fig:multidegree} we report the average normalised
multidegree of each of the 10 industry sectors listed in the ICB
classification \cite{ICB}.  We focus on the edges existing exclusively
in one of the four layers and on the combination of multi-edges
associated to edges existing in either of the Kendall, Tail, or
Partial layer, but not in the Pearson layer.
As shown in Fig.~\ref{fig:multidegree}, the multidegree exhibits
strong variations in time and high heterogeneity across different
industries.  Industries such as Oil $\&$ Gas, Utilities, and Basic
Materials, show low values of normalised multidegree in all the four
panels (a)-(d). Conversely, the edges of nodes corresponding to
Industrials, Finance, Technology, Telecommunications, and Consumer
Services tend to concentrate in one or in a small subset of layers
only.  For instance, we observe a relatively high concentration of
edges at the Kendall layer for nodes corresponding to Finance,
Industrials and Consumer Goods stocks in the period preceding the
Dot-com bubble and the 2002 downturn, a feature visible in the Pearson
layer in Fig.~\ref{fig:multidegree}(a).  Analogously, we notice a
sudden increase of edges unique to the Tail layer for nodes in
Consumer Goods, Consumer Services and Health Care after the 2007-2008
crisis. The presence of large heterogeneity and temporal variations in
the relative role of different industrial sectors confirms the
importance of using a multiplex network approach to analyse dependence
among assets.

From this perspective it is particularly interesting to discuss the
plot of multidegree restricted to edges that are present on either
Kendall, Partial or Tail layer, but are not present in the Pearson
layer as reported in Fig.~\ref{fig:multidegree}(e). Despite the
Pearson correlation coefficient is the most used measure to study
dependencies, the plot reveals that until 2002 an analysis of the
financial network based exclusively on Pearson correlations would have
missed from 40$\%$ up to 60$\%$ of the edges of assets in sectors such
as Basic Materials, Financial, Consumer Goods and Industrials.  The
study of evolution with time in Fig.~\ref{fig:multidegree}(e) reveals
that the relative role of such industrial sectors in Kendall, Tail and
Partial layers becomes relatively less important between the two
crises in 2002 and in 2007, but then such sectors become central again
during the 2007-2008 crisis and beyond.  This prominent role is quite
revealing but it would not had been evident from the analysis of the
Pearson layer alone.  Let us also note that, the period following the
2007-2008 crisis is also characterised by a sensible and unprecedented
increase of the normalised multidegree on Kendall, Partial and Tail
layers of stocks belonging to Technology and Telecommunications
sectors, whose importance in the market dependence structure has been
therefore somehow underestimated over the last ten years by the
studies based exclusively on Pearson correlation.

\subsection*{Multiplex cartography of financial systems}

\begin{figure*}[ht!]
 
  \includegraphics[width=2in]{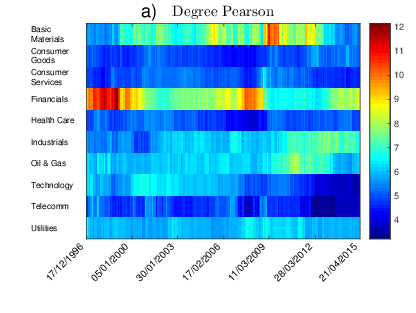}
  \includegraphics[width=2in]{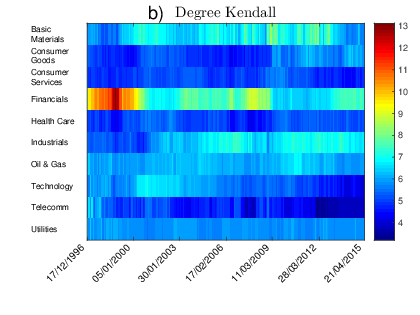}
  \includegraphics[width=2in]{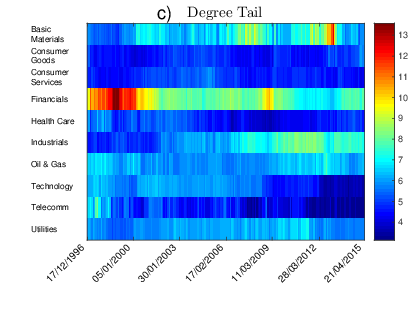}\\
  \includegraphics[width=2in]{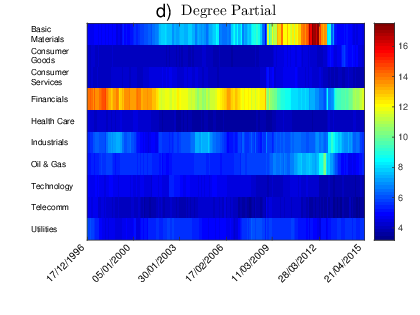}
  \includegraphics[width=2in]{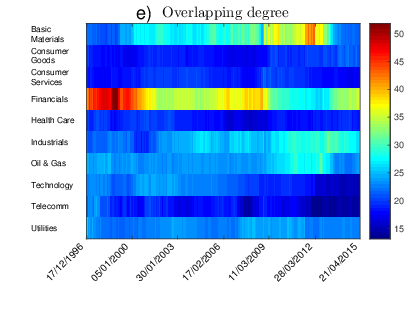}
  \includegraphics[width=2in]{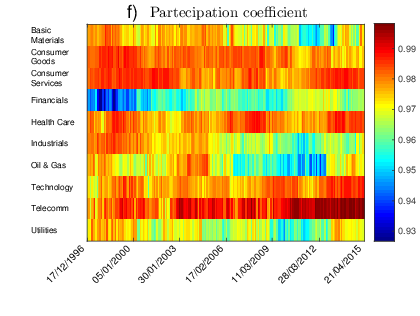}
\caption{\label{fig:overlap_degree} {\textbf{Average node degree as a
      proxy of the importance of an industry}}.  The plots of average
  degree of the nodes belonging to the different industrial sectors
  restricted to the (a) Pearson, (b) Kendall, (c) Tail, and (d)
  Partial layers, and of the average overlapping degree reported in
  panel (e) confirm the relative importance of Financials. However,
  the average participation coefficient (panel (f)) suggests that the
  dependence structure of some sectors such as Basic Materials,
  Industrials, and Oil \& Gas, has become more heterogeneous, i.e.,
  focused only on a subset of the four layers, after the 2007-2008
  crisis.}
\end{figure*}

To better quantify the relative importance of specific nodes and
groups of nodes we computed the overlapping degree
and partecipation coefficient, respectively measuring the total
number of edges of a node and how such edges are distributed across the
layers (see Materials and Methods for details).
We started by computing the average degree $k_I\lay{\alpha}$
at layer $\alpha$ of nodes belonging to each ICB industry sector $I$,
defined as 
$k_I\lay{\alpha}=\frac{1}{N_I}\sum_{i\in I} k_i\lay{\alpha}\delta(c_i, I)$,
where by $c_i$ we denote the industry of node $i$ and $N_I$ is the
number of nodes belonging to industry sector $I$. 
Figs. \ref{fig:overlap_degree} a)-d) show $k_I\lay{\alpha}$ as a
function of time for each of the four layers. 
%
Notice that nodes in the Financial sector exhibit a quite high average
degree, no matter the dependence measure used, with a noticeable peak
before the Dot-com bubble in 2002. After that, the average degree of
Financials has dropped sensibly, with the exception of the 2007-2008
crisis. Apart from the existence of similarities in the overall trend
of Financials across the four layers, the analysis of the average
degree suggests again the presence of high heterogeneity across
sectors and over time.

In the Pearson layer, Basic Materials is the second most central
industry throughout most of the observation interval, whereas
Industrials and Oil $\&$ Gas acquired more connections in the period
following the 2007-2008 crisis. The degree in the Kendall layer is
distributed more homogeneously among the sectors than in the Pearson
layer. Interestingly, the plot of degree on the Tail layer looks
similar to that of the Pearson layer. Finally, in the Partial layer we
observe the highest level of concentration of links in Finance
(consistently to what found in~\cite{partial_corr}) and, after the
2007-08 crisis, in Basic Materials.

We have also calculated for each industry $I$ the average overlapping
degree $o_I \equiv \langle o_i \rangle_{i \in I}$, where $o_i$ is the
overlapping degree of node $i$, which quantifies the overall
importance of each industrial sector in the multiplex dependence
network.  The average overlapping degree of each industry is shown as
a function of time in Fig.~\ref{fig:overlap_degree}(e). As we can see,
$o_I$ is able to highlight the prominent role played in the multiplex
network by Financials, Basic Materials, Oil \& Gas, and Industrials
sectors, revealing also the presence of four different phases between
1997 and 2015. The first phase, during which Financials is the only
prominent industry, covers the period between 1997 and 2000. The
second phase, between 2000 and the 2007-08 crisis, is characterised by
the emergence of Basic Materials as the second most central sector.
In the third phase, between 2009 and 2014, Financials  looses
its importance in favour of Industrials, Oil \& Gas and Basic
Materials (that becomes the most central one). Finally, in 2014 a new
equilibrium starts to emerge, with Financials and Industrials gaining
again a central role in the system.
   
The participation coefficient complements the information provided by
the overlapping degree, quantifying how the edges of a node are
distributed over the layers of the multiplex. In particular, the
participation coefficient of node $i$ is equal to $0$ if $i$ has edges
in only one of the layers, while it is maximum and equal to $1$ when
the edges of node $i$ are equally distributed across the layers (see
Materials and Methods for details). In
Fig.~\ref{fig:overlap_degree}(f) we report, as a function of time, the
average participation coefficient $P_I$ for each ICB industry
$I$. Interestingly, the plot reveals that the increase of the
overlapping degrees of Financials, Basic Materials, Industrials, and
Oil \& Gas sectors shown in Fig.~\ref{fig:overlap_degree}(e) are
normally accompanied by a substantial decrease of the corresponding
participation coefficients. This indicates that those sectors
accumulated degree on just one or two layers, confirming what we found
in multidegree analysis. A somehow more detailed analysis of the
temporal evolution of participation coefficient for each sector is
reported in Appendix.

\section*{Discussion}

By using filtered networks from different correlation measures we have
demonstrated that a multiplex network approach can reveal features
that would have otherwise been invisible to the analysis of each
dependency measure in isolation. Although the layers produced
respectively from Pearson, Kendall, Tail and Partial correlations show
a certain overall similarity, they exhibit distinct features that are
associated with market changes.  For instance, we observed that
average edge overlap between the first three layers, drops
significantly during periods of market stress revealing that
non-linear effects are more relevant during crisis periods.  The
analysis of the average multidegree associated to edges not present on
the Pearson layer, but existing on at least one of the three remaining
layers, indicates that Pearson correlations alone can miss to detect
some important features.  We observed that the relative importance of
non-linearity and tails on market dependence structure, as measured by
mean edge overlap between the last three layers, has dropped
significantly in the first half of 2000s and then risen steeply
between 2005 and the 2007-08 crisis. Overall, financial crises trigger
remarkable drops in the edge overlap, widening therefore the
differences among the measures of dependence just when evaluation of
risk becomes of the highest importance.
Different industry sectors exhibit different structural overlaps. For
instance, Financials, Industrials and Consumer Goods show an
increasing number of connections only on Kendall layer in the late
90s/early 2000, at the edge of the Dot-Com bubble.  After the 2007-08
crisis these industries tend to have many edges on the Kendall, Tail
and Partial which are not present on the Pearson layer. This
observation questions whether we can rely on the Pearson estimator
alone, when analysing correlations between stocks.
A study of the overlapping degree and of the participation coefficient
shows that asset centrality, an important feature for portfolio
optimization \cite{invest_periph,eccentricity_asset}, changes
considerably across layers with largest desynchronized changes
occurring during periods of market distress.
Overall our analysis indicates that different dependency measures
provide complementary informations about the structure and evolution
of markets, and that a multiplex network approach can reveal useful to
capture systemic properties that would otherwise go unnoticed.

\section*{Materials and Methods}

\subsection*{Data set}
\begin{figure}[ht!]
  \includegraphics[width=2.in]{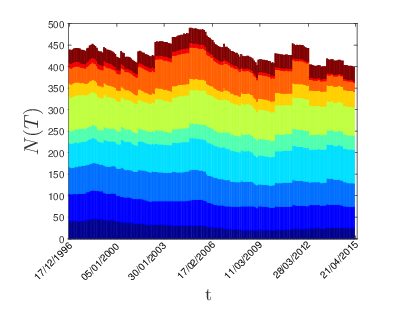}
  \includegraphics[width=1.3in]{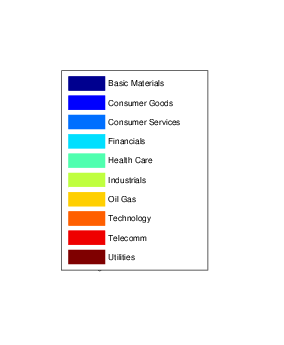}
  \caption{\label{fig:num_stocks} {\bf Number of stocks in each ICB
      industry in time}.  Number of stocks that are continuously
    traded in each time window together with their partition in terms
    of ICB industries.}
\end{figure}

The original dataset consists of the daily prices of $N_{tot}=1004$ US stocks 
traded in the period between 03/01/1993 and 26/02/2015. Each stock in
the dataset has been included in S\&P500 at least once
in the period considered. Hence the stocks considered 
provide a representative picture of the US stock market
over an extended time window of 22 years, and cover 
all the 10 industries listed in the Industry Classification
Benchmark (ICB) \cite{ICB} (Fig.\ref{fig:num_stocks}). 
It is important to notice that most of the 
stocks in this set are not traded over the entire period. This is a
major difference with respect to the majority of the works on dynamic
correlation-based networks, in which only stocks continuously traded
over the period under study are considered, leading to a significant
``survival bias''. For each asset $i$ we have calculated the series of
log-returns, defined as $r_i(t) = log(P_i(t)) - log(P_i(t-1))$, where $P_i(t)$
is stock price at day $t$.
The construction of the time-varying multiplex networks is based on log-returns and 
has been performed in moving time windows of $\theta = 1000$ trading days (about 4 years),
with a shift of $dT=23$ trading days (about one month), adding up to 200 different multiplex
networks, one for each time window. For each time window $T$, four different $N(T) \times N(T)$
dependence matrices have been computed, respectively based on the four different
estimators illustrated below. Since the number of active stocks changes with time,
dependence matrices at different times can have different number of
stocks $N(T)$, as shown in Fig.\ref{fig:num_stocks}. In the figure is also
shown the ICB industry composition of our dataset in each time window,
confirming that we have a representative sample of all market
throughout the period. We have verified that the results we are
discussing in the following are robust against change of $\theta$ and
$dT$.

\subsection*{Dependence among financial time series}

We have considered four different measures of dependence between two time series $r_i(u)$ and $r_j(u)$,
$i, j =1,2,\ldots,N$, $u=1,2,\ldots, \theta$, indicated
in the following respectively as Pearson, Kendall, Tail and Partial.

{\bf -- Pearson dependence --}~~
It is a measure of linear dependence between two time series, and is based on the evaluation of the Pearson correlation coefficient \cite{pearson}. We have used the exponentially smoothed version of this estimator \cite{exp_smoothing}, in order to  mitigate excessive sensitiveness to outliers in remote observations:

\begin{equation}
 \rho^w_{ij} =  \frac{\sum_{u=1}^{\theta} w_u(r_i(u) - \bar{r_i}^w)(r_j(u)-\bar{r_j}^w)}{\sqrt{\sum_{u=1}^{\theta} w_u(r_i(u)-\bar{r_i}^w)^2}\sqrt{\sum_{u=1}^{\theta} w_u(r_j(u)-\bar{r_j}^w)^2}} ~~,
\end{equation}
with 
\begin{equation}
w_u = w_0 \exp\left(\frac{u-\theta}{T^*} \right) ~~,
\end{equation}
where $T^*$ is the weight characteristic time ($T^*>0$) that controls the rate at which past observations lose importance in the correlation, and $w_0$ is a constant connected to the normalisation constraint $\sum_{u=1}^{\theta} w_u = 1$. We have chosen $T^* = \theta / 3$ according to previously established criteria \cite{exp_smoothing}. 
\\ 
\textbf{-- Kendall dependence --} ~~
It is a measure of dependence that takes into account the nonlinearity of a time series. It is based on the 
evaluation of the so-called Kendall's $\tau$ rank correlation coefficient, starting from the quantities 
$d_k(u,v) \equiv sgn(r_k(u) - r_k(v))$. The
estimator counts the number of concordant pairs, i.e. pairs of
observations such that $d_i(u,v)$ and $d_j(u,v)$ have equal signs, 
minus the number of discordant pairs \cite{kendall}. 
As for the case of the Pearson dependence, we have used the exponentially smoothed version of the 
estimator \cite{exp_smoothing}: 
\begin{equation}\label{kendall_weighted}
 \tau^w_{ij} = \sum^{\theta}_{u=1} \sum^{\theta}_{v=u+1} w_{u,v} ~d_i(u,v) d_j (u,v) ~~,
\end{equation}
with 
\begin{equation}
w_{u,v} = w_0 \exp\left(\frac{u-\theta}{T^*} \right) \exp\left(\frac{v-\theta}{T^*} \right) ~~,
\end{equation}
where $T^*$ is again the weight characteristic time.\\
\textbf{-- Tail dependence --} ~~ It is a non-parametric estimator of tail copula that  
  provides a measure of dependence focused on extreme events. It is based on the 
evaluation of the following estimator \cite{tail_corr}: 
  \begin{equation}
 C_{ij}(p_1,p_2) = \frac{\sum_{u=1}^{\theta} \mathbbm{1}_{ \{F^i(r_i(u))< p_1 \wedge  F^j(r_j(u))< p_2 \}}}{\sum_{u=1}^{\theta} \mathbbm{1}_{ \{F^i(r_i(u))< p_1 \vee  F^j(r_j(u))< p_2 \}}}
 \end{equation}
 where $F^i$ and $F^j$ are the empirical cumulative probabilities of
 returns $r_i(u)$ and $r_j(u)$ respectively, $p_1$ and $p_2$ are two
 parameters representing the percentiles above which an observation is
 considered (lower) tail.  We focus on lower tails since we are
 interested in risk management applications, where the attention is on
 losses. It can be shown that this is a consistent estimator of tail
 copula \cite{tail_corr}. In this work we have chosen $p_1=p_2=0.1$
 (i.e. we consider tail every observation below the 10th percentile),
 as a trade-off between the need of statistic and the interest in
 extreme events.
\\ 
\textbf{-- Partial dependence--}~~ 
 It is a measure of dependence that quantifies to what extent each asset affects other
 assets correlation. The Partial correlation $\rho_{ik|j}$, or correlation influence, 
 between assets $i$ and $k$ based on $j$, is the Pearson correlation between
 the residuals of $r_i(u)$ and $r_k(u)$ obtained after regression
 against $r_j(u)$ \cite{partial_corr2}. It can be written in terms of a 
 Pearson correlation coefficient as follows \cite{partial_corr}:
 \begin{equation}
  \rho_{ik|j} = \frac{\rho_{ik} - \rho_{ij} \rho_{kj}}{\sqrt{[1-\rho_{ij}^2][1-\rho_{kj}^2]}}
 \end{equation}
This measure represents the amount of correlation between $i$ and $k$
that is left once the influence of $j$ is subtracted. Following
\cite{partial_corr}, we define the correlation influence of $j$ on the
pair $i,k$ as:
 \begin{equation}
  d(i,k|j)= \rho_{ik} - \rho_{ik|j}
 \end{equation}
$d(i,k|j)$ is large when a significant fraction of correlation
 between $i$ and $k$ is due to the influence of $j$. Finally, in order
 to translate this into a measure between $i$ and $j$, the so-called Partial dependence, 
 we average it over the index $k$:
 \begin{equation}
  d(i|j) = \langle d(i,k|j) \rangle_{k \neq i,j}
 \end{equation}
$d(i|j)$ is the measure of influence of $j$ on $i$ based on Partial
 correlation. It is worth noting that, unlike the other measures of dependence,
 $d(i|j)$ provides a directed relation between assets (as in general $d(i|j) \neq d(j|i)$). In the
 rest of the paper we refer to this indicator as ``Partial dependence'', even
 though strictly speaking we are analysing the Correlation influence
 based on Partial correlation.

\subsection*{Graph filtering and the construction of the multiplex network}

For each of the 200 time windows we have then constructed a multiplex network with
$M=4$ layers obtained respectively by means of the four dependence indicators. 
In order to reduce the noise and the redundance contained in each
dependence matrix we have applied the Planar Maximally Filtered Graph
\cite{Tumminello05} \cite{PMFG2} \cite{PMFG3} \cite{TMFG}.
It is worth mentioning that the filtering of the correlation influence layer
requires an adaptation of the PMFG algorithm  to deal with
asymmetric relations. We have followed the approach suggested in \cite{partial_corr}
that rules out double links between nodes. The obtained planar graphs have been then
converted into an undirected graphs and included in the multiplex.

\subsection*{Multiplex measures}

Let us consider a weighted multiplex network $\mathcal{M}$ on $N$
nodes, defined by the $M$-dimensional array of weighted adjacency
matrices $\mathcal{W} = \{W\lay{1}, W\lay{2}, \ldots, W\lay{M}\}$,
where $W\lay{\alpha}=\{w_{ij}\lay{\alpha}\}$ are the matrices of
weights that determine the topology of the $\alpha$-th layer though
the PMFG filtering. Here the weight $w_{ij}\lay{\alpha}$ represents
the strength of the correlation between node $i$ and node $j$ on layer
$\alpha$, where the different layers are obtained through different
correlation measures. In the following we will indicate by
$W\lay{\alpha}$ the weighted adjacency matrix of the PMFG associated
to layer $\alpha$, and by $A\lay{\alpha}$ the corresponding unweighted
adjacency matrix, where $a_{ij}\lay{\alpha} = 1$ if and only if
$w_{ij}\lay{\alpha} \not= 0$. We denote by
$K\lay{\alpha}=\frac{1}{2}\sum_{ij}a_{ij}\lay{\alpha}$ the number of
edges on layer $\alpha$, and by $K = \frac{1}{2}\sum_{i,j} \left[1 -
  \prod_{\alpha}(1-a_{ij}\lay{\alpha})\right]$ the number of pairs of
nodes which are connected by at least one edge on at least one of the
$M$ layers. Notice that since the network at each layer is a PMFG,
then we have $K\lay{\alpha} = 3(N-2)\> \forall \alpha$ by
construction.

We  consider some basic quantities commonly used to characterise
multiplex networks~\cite{Bianconi2013,metrics_multiplex}. The first
one is the mean edge overlap, defined as the average number of layers
on which an edge between two randomly chosen nodes $i$ and $j$ exists:
\begin{equation}
  \avg{O} = \frac{1}{2K}\sum_{i,j} \sum_{\alpha}
  a_{ij}\lay{\alpha}.
\end{equation}
Notice that $\avg{O}=1$ only when all the $M$ layers are identical,
i.e. $A\lay{\alpha}\equiv A\lay{\beta}\> \forall \alpha,
\beta=1,\ldots,M$, while $\avg{O}=0$ if no edge is present in more
than one layer, so that the average edge overlap is in fact a measure
of how much similar is the structure of the layers of a multiplex
network.
A somehow dual quantity is the fraction of edges of layer $\alpha$
which do not exist on any other layer:
\begin{equation}
  U\lay{\alpha} = \frac{1}{2K\lay{\alpha}}\sum_{i,j}a_{ij}\lay{\alpha}
  \prod_{\beta\neq \alpha} \left(1-a_{ij}\lay{\beta}\right)
\end{equation}
which quantifies how peculiar is the structure of a given layer
$\alpha$, since $U\lay{\alpha}$ is close to zero only when almost all
the edges of layer $\alpha$ are also present on at least one of the
other $M-1$ layers. 

More accurate information about the contribution of each node to a
layer (or to a group of layers) can be obtained by the so-called
multidegree of a node $i$. Let us consider the vector $\vec m =
(m_1,m_2,...,m_M)$, with $M$ equal to the number of layers, where each
$m_{\alpha}$ can take only two values $\{1,0\}$. We say that a pair of
nodes $i,j$ has a multilink $\vec m$ if they are connected only on
those layers $\alpha$ for which $m_{\alpha}=1$ in $\vec
m$~\cite{Bianconi2013}.  The information on the $M$ adjacency matrices
$a_{ij}^{\alpha}$ ($\alpha=1,..,M$) can then be aggregated in the
multiadjacency matrix $A_{ij}^{\vec m}$, where $A_{ij}^{\vec m}=1$ if
and only if the pair $i,j$ is connected by a multilink $\vec
m$. Formally~\cite{Bianconi2013,multiplex_review}:
 \begin{equation}
  A_{ij}^{\vec m} \equiv \prod_{\alpha=1}^M [a_{ij}^{\alpha}m_{\alpha}+(1-a_{ij}^{\alpha})(1-m_{\alpha})].
 \end{equation}
From the multiadjacency matrix we can define the multidegree $\vec m$
of a node $i$, as the number of multilinks $\vec m$ connecting $i$:
 \begin{equation}
  k_{i}^{\vec m} = \sum_{j} A_{ij}^{\vec m}.
 \end{equation}
This measure allows us to calculate e.g. how many edges node $i$ has
on layer $1$ only ($k_{i}^{\vec m}$ choosing $m_1=1$, $m_{\alpha}=0 ~~
\forall \alpha \neq 1$), integrating the global information provided
by $U^{[\alpha]}$.

The most basic measure to quantify the importance of single nodes on
each layer is by means of the node degree $k_i\lay{\alpha} =
\sum_{j}a_{ij}\lay{\alpha}$. However, since the same node $i$ is
normally present at all layers, we can introduce two quantities to
characterise the role of node $i$ in the
multiplex\cite{metrics_multiplex}, namely the overlapping degree
\begin{equation}
  o_i = \sum_{\alpha}k_i\lay{\alpha}
\end{equation}
and the multiplex participation coefficient:
\begin{equation}
  P_i = \frac{M}{M-1}\left[1 -
    \sum_{\alpha}\left(\frac{k_i\lay{\alpha}}{o_i}\right)\right].
\end{equation}
The overlapping degree is just the total number of edges incident on
node $i$ at any layer, so that node are classified as \textit{hubs} if
they habve a relatively large value of $o_i$. The multiplex
participation coefficient quantifies the dispersion of the edges
incident on node $i$ across the layers. In fact, $P_i=0$ if the edges
of $i$ are concentrated on exactly one of the $M$ layers (in this case
$i$ is a \textit{focused} node), while $P_i=1$ if the edges of $i$ are
uniformly distributed across the $M$ layers, i.e. when
$k_i\lay{\alpha} = \frac{o_i}{M}\> \forall \alpha$ (in which case $i$
is a \textit{truly multiplex} node). The scatter plot of $o_i$ and
$P_i$ is called \textit{multiplex cartography} and has been used as a
synthetic graphical representation of the overall heterogeneity of
node roles observed in a multiplex.

In a multiplex network is important also to look at the presence and sign of inter-layer
degree correlations. This can be done by computing the inter-layer degree correlation
coefficient~\cite{Nicosia2014correlations}:
\begin{equation}
  \rho\lay{\alpha, \beta} = \frac{\sum_{i}\left( R\lay{\alpha}_i -
    \overline{ R\lay{\alpha} } \right) \left( R\lay{\beta}_i -
    \overline{ R\lay{\beta} } \right)} {\sqrt{\sum_i\left(
      R\lay{\alpha}_i - \overline{ R\lay{\alpha} } \right)^2\sum_j
      \left( R\lay{\beta}_j - \overline{ R\lay{\beta} } \right)^2}}
\end{equation}
where $R_i\lay{\alpha}$ is the rank of node $i$ according to its
degree on layer $\alpha$ and $\overline{R\lay{\alpha}}$ is the average
rank by degree on layer $\alpha$. In general $\rho\lay{\alpha,\beta}$
takes values in $[-1,1]$, where values close to $+1$ (respectively,
$-1$) indicate the of strong positive (resp. negative) correlations,
while $\rho\lay{\alpha,\beta}\simeq 0$ if the degrees at the two layers
are uncorrelated.

\section*{APPENDIX}
\subsection*{Time evolution of the average participation coefficient}
\begin{figure*}[ht!]
  \includegraphics[scale=0.4]{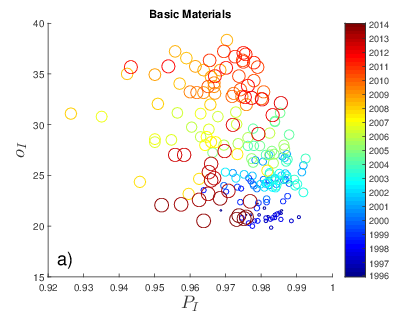}
  \includegraphics[scale=0.4]{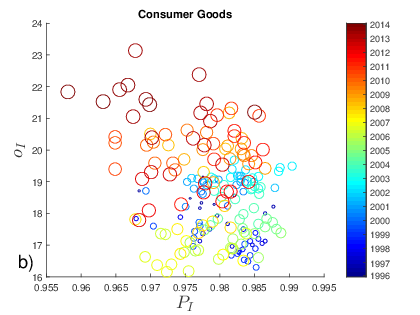}
  \includegraphics[scale=0.4]{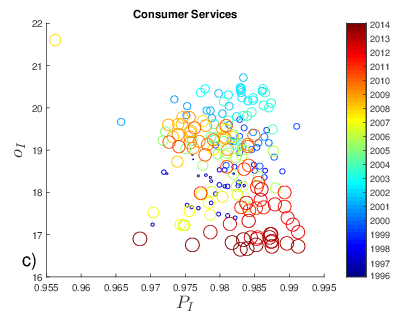}
  \includegraphics[scale=0.4]{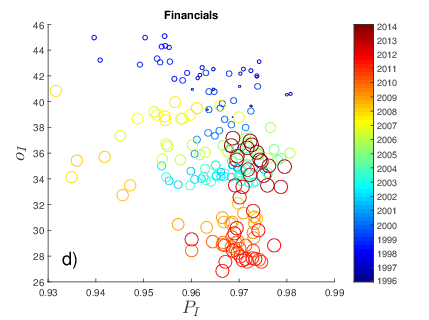}
  \includegraphics[scale=0.4]{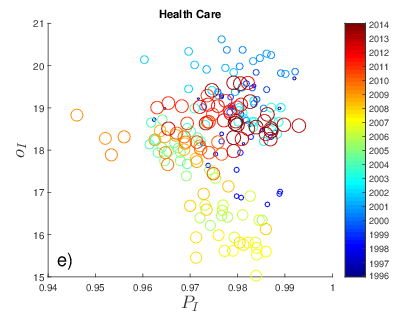}
  \includegraphics[scale=0.4]{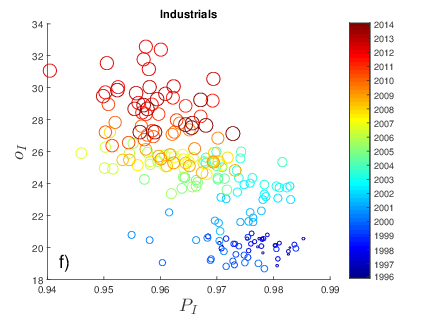}
	 \includegraphics[scale=0.4]{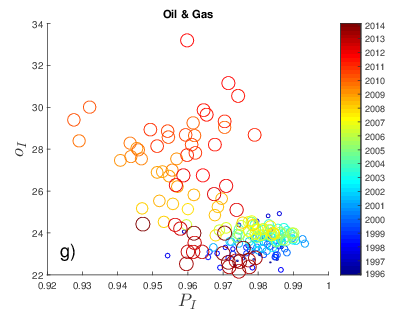}
  \includegraphics[scale=0.4]{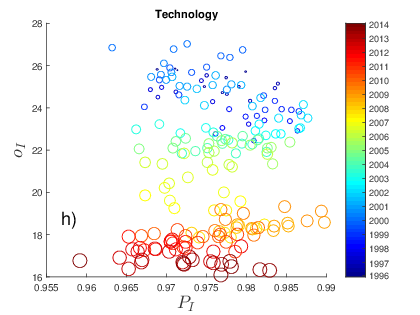}
  \includegraphics[scale=0.4]{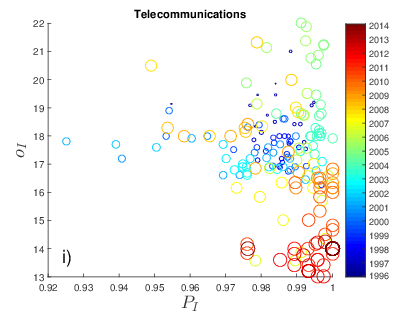}
  \includegraphics[scale=0.4]{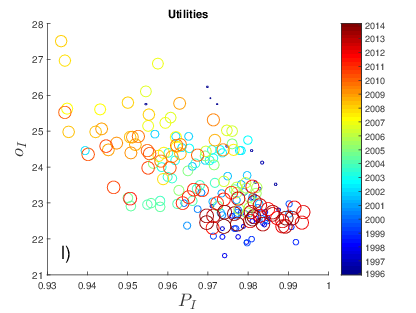}
  
\caption{\label{fig:cartography1} {\bf Industries evolution in the
    overlapping degree/partecipation coefficient plane}. Fixed an
  industry $I$, we have plotted for each time window a circle whose
  $y$ coordinate is the average overlapping degree $o_I$ and whose $x$
  coordinate is the average partecipation coefficient $P_I$.  Points
  at different times are characterized with different sizes (small to
  large) and colours (legend on the right). In a) - l) we show the
  results respectively for Basic Materials, Consumer Goods, Consumer
  Services, Financials, Health Care, Industrials, Oil \& Gas,
  Technology, Telecommunications and Utilities.}
\end{figure*}

In Fig.~\ref{fig:cartography1} we plot the time evolution of the
average participation coefficient $P_I$ (x-axis) of stocks in the
industrial sector $I$ against the average overlapping degree $o_I$
(y-axis). Each circle corresponds to one of the 200 time windows,
while the size and colour of each circles represent different time
windows. Each panel corresponds to one industrial sector $I$.
The diagrams reveal that in the last 20 years the role of different
sectors has changed radically, and in different directions. For
instance, stocks in the Financials sector evolved from a relatively
large overlapping degree and a small participation coefficient in the
late 1990s, to a smaller number of edges, distributed more
homogeneously across the layers, towards the end of the observation
period. Conversely, Industrials stocks have acquired degree on some of
the layers, resulting in a considerable decrease of participation
coefficient. This is another indication of the importance of
monitoring all the layers together, as an increase in the structural
role of an industry (as measured by the overlapping degree) is
typically due to only a subset of layers (as indicated by the
corresponding decrease of partecipation coefficient).

\section*{Acknowledgments}
The authors wish to thank Alessandro Fiasconaro for useful discussions
at the beginning of this project. V.L. acknowledges support from the
EPSRC project GALE, EP/K020633/1. The authors wish to thank Bloomberg
for providing the data. TDM wishes to thank the COST Action TD1210 for
partially supporting this work.

\section*{Author contributions statement}

N.M. and V. N. contributed equally to this work. All the authors
devised the study, performed the experiments and simulations, analysed
the results, wrote the paper, and approved the final draft.

\section*{Additional information}

\textbf{Competing financial interests.} The authors declare no
competing financial interest.

\end{document}